\documentclass[11pt]{article}
\usepackage{epsfig}

\usepackage[vmarginratio=1:1,hmarginratio=1:1,
totalwidth=16.5cm,totalheight=21.26cm]{geometry}
\def\baselinestretch{1.31}

\newcommand{\be}{\begin{equation}}
\newcommand{\ee}{\end{equation}}
\newcommand{\bea}{\begin{eqnarray}}
\newcommand{\eea}{\end{eqnarray}}

\DeclareMathSymbol{\mg}{\mathrel}{symbols}{"1D}

\newcommand{\ga}{\alpha}

\renewcommand{\ge}{\epsilon}

\newcommand{\gth}{\theta}

\newcommand{\gL}{\Lambda}

\newcommand{\cI}{{\cal I}}
\newcommand{\cJ}{{\cal J}}

\newcommand{\cM}{{\cal M}}
\newcommand{\cN}{{\cal N}}
\newcommand{\cO}{{\cal O}}

\newcommand{\bZ}{{\bf Z}}
\newcommand{\ra}{\rightarrow}

\newcounter{oldcounter}
\addtocounter{equation}{1}
\begin{document}
\begin{flushright} 
DAMTP-2002-140.
\end{flushright} 
\vskip 3cm
\begin{center} 
{\Large {\bf  On Gauge couplings, ``Large'' extra-dimensions\\
\vspace{0.4cm}
and the limit $\alpha'\ra 0$ of the String.\\}}
\bigskip 
\vspace{0.93cm} 
{
{\bf D.M. Ghilencea\footnote{
{{ {\ {\ {\ E-mail: D.M.Ghilencea@damtp.cam.ac.uk}}}}}}}}
\\
\vspace{0.9cm} 
{\it DAMTP, CMS, University of Cambridge} \\
{\it Wilberforce Road, Cambridge, CB3 0WA, United Kingdom.}\\
\bigskip 
\end{center}
\vspace{1.cm}
\begin{center}
{\bf Abstract}\\
%\vspace{0.5cm}
\end{center}
{
Using an effective field theory (EFT)  approach for a generic
model with two additional dimensions compactified on a two-torus, 
we compute the  total one-loop radiative corrections to the gauge 
couplings due to associated Kaluza-Klein massive  {\it and}  massless  
states.  A consistent treatment of  both infrared (IR) and ultraviolet 
(UV) divergences shows a connection via infrared regulator effects 
between the massless and massive  sectors   of a compactified
theory. A new correction to the gauge couplings is found such that   
their  UV behaviour is sensitive to IR regulator dependent effects in 
the  sector of (infinitely many)  massive  modes. This correction is 
a one-loop UV-IR mixing effect due to infinitely many Kaluza-Klein 
modes and exists for two compact dimensions.
The link with string theory is addressed to show that this correction, 
logarithmic in the UV scale,  cannot be recovered from known (infrared 
regularised)  string calculations  in the field theory  ``limit'' 
$\alpha'\ra 0$. We explain the origin of this discrepancy and address 
some of its implications.}

\newpage\setcounter{page}{1}

\section*{Introduction.}

String theory is usually thought of as a more fundamental theory
of particle physics than  current effective field theory
descriptions are.
The latter  are very successful in describing accurately many
aspects of particle physics, but it is thought that ultimately,
at some (very) high energy scale, a string description is supposed to 
provide a  fully consistent
 theoretical framework of high energy physics. In the low energy
limit, one would hope to recover the effective field theory 
description.  However, the precise link between these 
two theories is not always clear.  To clarify this
relationship, physicists try to reproduce previous 
string-derived results using effective field theory approaches.

It is the purpose of this work to address such a link for
the problem of radiative  corrections to the gauge couplings in the
context of an effective field theory model compactified on a 
two-torus. Although we seek to clarify the link with (heterotic) 
string theory, the calculation is also relevant  in the context of models
with ``large'' extra-dimensions, without reference to string theory.  
The model is assumed to  be a 4D $\cN\!=\! 1$ supersymmetric orbifold with 
an $\cN\!=\!2$ sub-sector  (e.g. $Z_4$ orbifold)
which ``survives'' after compactification. The two  extra dimensions 
compactified on the two-torus
have associated Kaluza-Klein  states -- and at string level winding
states as well --  which fall into $\cN=2$ multiplets,
charged under the gauge group  of the model. These
states bring   significant one-loop corrections 
to the gauge couplings.  At (heterotic) string level, such corrections
are well-known  \cite{Dixon:1990pc}  (see also  \cite{Mayr:1993mq}, 
\cite{Kaplunovsky:1995jw}, \cite{Mayr:1995rx}), 
and will also be the subject of the discussion below.

An  effective field theory (EFT) analysis of such one-loop
corrections can only account for the effects of momentum states.
Such analysis would correspond at  string level 
to the limit  of an infinite string scale or $\alpha'\ra 0$ 
(with $M_s^2\propto 1/\alpha'$ \cite{Kaplunovsky:1987rp}),
when winding modes' effects are suppressed. 
It was shown in  \cite{Ghilencea:2002ff}  that such an
EFT approach can recover the (gauge dependent part of the)  
string result for the 
radiative corrections to the  gauge couplings due to {\it massive} 
momentum and winding modes {\it only}, in the limit  $\alpha'\ra 0$ (and up 
to worldsheet instantons contribution, vanishing in 
this limit). This is important for the EFT approach does not respect
the full symmetries and spectrum of the string. For further details
on this EFT approach and model description see  \cite{Ghilencea:2002ff}. 
Here we continue this EFT  analysis, to  reveal an additional, 
interesting effect.

The one-loop corrections  to the gauge coupling $\alpha_i$ 
in a $\cN=1$  orbifold  are usually separated  as
\begin{equation}
\alpha _{i}^{-1}(Q) = \alpha _{o}^{-1}(M_s)+\frac{b_{i}}{4\pi } \ln
\frac{M_{s}^2}{Q^2}+\Omega_{i},\qquad\textrm{$i$: gauge group index}
\label{gauge}
\end{equation}
$Q$ is the low energy scale, $M_s$ is the string/UV  scale, $\alpha_o$
is the ``bare'' coupling, the ``log'' term is due to
$\cN\!=\! 1$ {\it massless} modes correction, usually identified 
with MSSM-like  spectrum, with $b_i$ one-loop beta function.  
$\Omega_i$ is the (gauge dependent part of the) one-loop
correction due to  {\it massive} $\cN\!=\! 2$ multiplets/states 
(momentum and in string case  winding states as well). 

A (heterotic) string calculation  \cite{Dixon:1990pc} of $\Omega_i$ gives
\begin{equation}\label{t2form}
\Omega_{i}^{H} =-\frac{\overline{b}_{i}}{4\pi }\ln \Big[ 
C_{reg.} \,U_2\,{|}\eta (U){|}^{4}\, 
T_{2}\left| \eta \left( {T}\right) \right|^{4}\Big], 
\qquad 
\eta(T)\equiv e^{\frac {\pi}{12} iT}
\prod_{k=1}^{\infty}\left( 1-e^{2\pi ikT} \right). 
\end{equation}
$\overline{b}_{i}$ is the beta function coefficient 
in the ${\cal N}\!=\!2$ sub-sector, $C_{reg.}$ is an {\it infrared}
regularisation constant. $T = T_1 + i T_2$, 
$U = U_1 + i U_2$ are moduli expressed in function of the metric $G_{ij}$
(of determinant $G$) and anti-symmetric tensor background $B_{ij} = B
\ge_{ij}$ as 
$T = 2 [ B+ i \sqrt G ] = 
2 [ B + i {R_1 R_2 \sin \gth}/{(2 \ga')} ],$
%%%\qquad 
and $U = {1}/{G_{11}}[ G_{12}+i \sqrt G ] 
= {R_2}/{R_1} \exp(i \gth).$ 
$T_2$ is expressed in $\alpha'$ units and 
describes the area of compactification
 while $U$ describes the shape of the torus ($\ga'$ independent).

The EFT result for $\Omega_i$, of summing one-loop effects of {\it massive} 
Kaluza-Klein modes only 
\cite{Ghilencea:2002ff} is 
\begin{equation}\label{result}
\Omega_{i}^{EFT}= - \frac{{\overline {b}}_i}{4 \pi} 
\ln\Big[C_{reg.}  \, U_2\, \vert \eta(U)\vert^4 \, e^{-{T_2^*}} \, 
{T_2^*}\Big], 
\qquad T_2^*\equiv  \Lambda^2  R_1 R_2 \sin\theta; \qquad
\textrm{$\Lambda$: UV scale}
\end{equation}
Up to a UV regulator 
re-scaling/re-definition\footnote{This 
re-scaling by a constant $(3/\pi)$ is ultimately
\cite{Ghilencea:2002ff} an effect of winding modes
 (modular invariance) and has no physical meaning at EFT level; it
 only relates the EFT regulator to modular invariance
 ``regularisation''.} \cite{Ghilencea:2002ff},
eq.(\ref{result})  has identical (UV divergent) behaviour to 
the limit  $\alpha'\ra 0$  of the string result (\ref{t2form}) 
which includes the effects of the winding modes as well. 
If one uses the same regularisation in both cases,
the regularisation constants $C_{reg.}$ in (\ref{t2form}) and
(\ref{result}) are equal, but unlike the string case, 
$C_{reg.}$  of (\ref{result}) is due  to an UV regularisation (not  IR
as in eq.(\ref{t2form})).

In  the  analysis below  we  address further subtleties
of the gauge couplings corrections in this two-torus
compactification. So far we mentioned the one-loop effects of the {\it
massive} (momentum, and in string case also winding)
 modes on the gauge couplings. Orbifold models also have
 massless modes, for example (but not only) the
twisted modes  ``living'' at the fixed points of the 
orbifold\footnote{Other massless states exist, for example if 
Wilson lines are present. The effect  discussed  is not 
(necessarily) due to (massless) twisted states, but more generally
to the existence of massless states in a infinite Kaluza-Klein tower.}.
Considered separately, the  massless states bring a  logarithmic
correction as  in eq.(\ref{gauge}), in addition to the massive 
modes' correction. In this work we  analyse more closely effects 
related to these states.

To compute the  massless modes corrections to $\alpha_i$ 
one needs  an IR regularisation.  An UV regularisation is
also required for their logarithmic  correction.  Surprisingly, 
the IR regularisation  may have implications  for the radiative
effects 
of the massive  Kaluza-Klein modes as well, in the  EFT
approach that we adopt in this work. Apparently it would seem 
that these two mass sectors are not connected. However, 
  consistency of the regularisation requires us keep an IR 
regulator $\chi$  in the  massive sector as well (only UV
divergent in the EFT case), and only in  the end remove this
regulator, $\chi\ra 0$. The result is the presence of an  EFT
correction to the gauge couplings which is 
 a ``mixing'' term $\chi \ln \xi$ with  UV regulator $\xi\ra 0$ as well.
 This brings a ``non-decoupling'' of the IR effects 
from a  UV logarithmically {\it divergent}  correction. This correction
must be added in the r.h.s. of eq.(\ref{gauge}). This is a
combined effect of the massless and massive sectors (via infrared
effects).

This correction  is not recovered from string calculations.
In  the one-loop string calculation of $\Omega_i^H$ (only) 
an IR regularisation is needed  \cite{Dixon:1990pc}.
The aforementioned  EFT correction can have a string  correspondent 
of type $\epsilon\ln\alpha'$, with $\epsilon\ra 0$ acting now as IR 
string regulator. As we will discuss, such terms do appear in string 
calculations.
This term vanishes in the limit $\epsilon\!\ra\! 0$ because
$\alpha'$ is  {\it non-zero} and the limit 
of removing the IR string regulator  in such terms is 
then  possible/allowed. If we then take the field theory ``limit''
 $\alpha'\ra 0$ on the string, terms like $\epsilon\ln\alpha'$
are not encountered/kept.  Thus the limit $\alpha'\ra 0$ on the
(regularised) string result will  not recover the 
aforementioned EFT correction. This seems generic for the  limit 
$\alpha'\ra 0$  of string calculations with infrared regularisation.

String symmetries (e.g.  modular invariance)
ensure that calculations done in this framework
show a better (finite) UV behaviour than the EFT does, though
 the string  does require IR regularisation. When
$\alpha'\!\ra\! 0$ such symmetries do not necessarily survive, UV
divergences (in $\alpha'$) appear, may ``interfere'' 
with the IR regularisation and string symmetries (now broken)
cannot teach us anymore in which order to remove the IR/UV regulators.
 If correct, the presence of the new EFT correction 
also tells us that one cannot address  the UV behaviour 
(the $\!\alpha'\!\ra\! 0$ limit) of a string
 model without simultaneously addressing its IR behaviour and 
the limit  $\alpha'\ra 0$  of the (regularised) string may
 not always  recover the  UV behaviour 
as found in the corresponding EFT theory!

In the next section one-loop corrections are computed in the EFT
approach with particular attention to the effect of massless states.
The link with string theory and its $\alpha'\ra 0$ limit is then
discussed, followed by  conclusions. Some mathematical
details  are presented in the Appendix.

%%%%%%%%%%%%%%%%%%%%%%%%%%%%%%%%%%%%%%%%%%%%%%%%%%%%%%%%%%%%%%%

\section*{One-loop corrections to the gauge couplings.}

%%%%%%%%%%%%%%%%%%%%%%%%%%%%%%%%%%%%%%%%%%%%%%%%%%%%%%%%%%%%%%%

To give quantitative support to our statements, 
we use the Coleman-Weinberg formula for 
one-loop radiative effects to the gauge couplings, 
see for example \cite{Kaplunovsky:1987rp}.  The one-loop corrected 
coupling is 
 \begin{equation}\label{totalcorrection}
\alpha_i^{-1}\bigg\vert_{{one-loop}}=
\alpha_i^{-1}\bigg\vert_{{tree-level}}+\Omega_i^T,\qquad 
i: \textrm{gauge group index}
\end{equation}
In the context of two extra dimensions
compactified on a two-torus the Coleman-Weinberg 
expression for  $\Omega_i^T$ due to summing the effects of all
states in theory is 
\begin{equation}\label{QFT0}
\Omega_{i}^T = \frac{1}{4\pi}\sum_{R} T_i(R) 
\sum_{m_{1,2} \in \bZ} \int_{0}^{\infty}\frac{dt}{t} 
\, e^{-\pi\,t M_{m_1,m_2}^2/\mu^2 }\bigg\vert_{reg.}
\end{equation}
$\mu$ is a finite, non-zero mass parameter introduced to 
set this equation dimensionless. $T_i(R)$ is the beta function
contribution of a state - charged under the symmetry group of the
model - and which has an associated Kaluza-Klein tower.
The sum over the integers $m_{1,2}$ runs over all Kaluza-Klein 
states of mass $M_{m_1,m_2}$
and the subscript ``reg'' emphasizes that the above expression 
only makes sense in the presence of a regularisation. 
Indeed, integral (\ref{QFT0})  is divergent in the UV ($t\ra 0$) so
 a regulator $\xi$ is introduced ($\xi\ra 0$).  For the massless 
states/modes 
the above integral is also IR divergent ($t\ra \infty$), so an 
IR regulator $\chi$ ($\chi\ra 0$)  is also required. A well
defined formula is then\footnote{Other regularisations are possible. For a
discussion on this see \cite{Ghilencea:2002ff}.}
\begin{equation}\label{QFTthresholds0}
\Omega_{i}^T = \frac{1}{4\pi}\sum_{R} T_i(R) 
\sum_{m_{1,2} \in \bZ} \int_{\xi}^{\infty}\frac{dt}{t} 
\, e^{-\pi\,t M_{m_1,m_2}^2/\mu^2 } e^{-\pi \chi t}
\end{equation}
To evaluate the above integral we must specify the compactification
manifold/spectrum. To keep the analysis model independent, we note
that  for a  two-torus compactification 
(radii $R_{1,2}$, cycles' angle $\theta$, $\theta=\pi/2$ for
orthogonal torus) the masses of Kaluza-Klein states  have the  
generic structure (see details in  e.g. \cite{Dienes:2001wu}) 
\begin{equation}\label{kkmass}
M^2_{m_1,m_2}=
\frac{1}{\sin^2\theta}\left[\,\frac{m_1^2}{R_1^2}+\frac{m_2^2}{R_2^2}-
\frac{2 m_1 m_2 \cos\theta}{R_1 R_2}\,\right]
\,\, =\,\, \frac{\vert m_2-U m_1\vert^2 }{ (\mu^{-2} T_2) U_2} .
\end{equation}
In the EFT approach we use the following definitions
for  the (dimensionless) quantities  $T$ and $U$
\begin{equation}\label{fieldcase}
{T} (\mu) \equiv i\, T_2 (\mu)= i\, \mu^2 R_1 R_2 \sin\theta,\qquad
U=U_1+i\, U_2=\frac{R_2}{R_1} \, e^{i \theta}
\end{equation}
In eq.(\ref{QFTthresholds0}) one can isolate the contribution of the 
massless mode (in our case  $(0,0)$) from that of massive modes,
in our case  $(m_1,m_2)\not=(0,0)$ as
\begin{equation}\label{QFT1}
\Omega_{i}^T = \Omega_i^*+\Omega_i^{0}
\end{equation}
where\footnote{We denote by $\gamma$  the Euler constant, $\gamma=0.57721$}
\begin{eqnarray}
\Omega_i^{0}&\equiv &\frac{b_i}{4 \pi} \int_{\xi}^{\infty}
\frac{dt}{t} e^{- \pi \chi t} =\frac{b_i}{4 \pi} \Gamma(0,\pi \chi \xi)
=\frac{b_i}{4 \pi}
\ln \frac{\,\, 1/\xi}{ (\pi e^{\gamma}) \, \chi} \label{QFT2}\\
\nonumber\\
\Omega_i^* &\equiv & \frac{{\overline b_i}}{4\pi}
\sum_{m_{1,2} \in \bZ}' \int_{\xi}^{\infty}\frac{dt}{t} 
\, e^{-\pi\,t M_{m_1,m_2}^2/\mu^2 }  e^{-\pi \chi t}\label{QFT3}
\end{eqnarray}
A ``prime'' on the double sum stands for the 
absence of $(0,0)$ mode.
 $\Omega_i^0$ is due to massless modes,
and $\Omega_i^*$ is due to massive modes only.
In general $\Omega_i^0$ and $\Omega_i^*$ have in front
different beta functions coefficients ($b_i$, $\overline b_i$). 
The exact value of the latter,
a sum (of $T_i(R)$) over a model dependent spectrum, 
is not relevant for our purposes. In general $b_i\not=\overline b_i$
since in  the class of models we  address  the massless 
modes include ``twisted'' states (MSSM-like matter states)
 living at the fixed points  of the orbifold considered, where  
the spectrum is different 
from that of the ``bulk'' modes  $(m_1,m_2)\!\!\not=\!\!(0,0)$.

$\Omega_i^0$ is both IR and UV (logarithmically)
 divergent as the regulators dependence shows. Ultimately the
 logarithm in (\ref{QFT2}) is just the log term  of 
(\ref{gauge}), as we will see upon relating the  regulators
($\chi$ and $\xi$) to some IR and UV physical scales ($Q$ and $\Lambda$)
 respectively. We now address the massive/''bulk''
  modes' correction $\Omega_i^*$.  Except its $\chi$ additional 
dependence,  $\Omega_i^*$ is identical  to that in (\ref{result}) and in
\cite{Ghilencea:2002ff},  where the effect of massive modes alone
was analysed. Since $\Omega_i^*$ is  IR finite\footnote{If Wilson
lines exist, ``bulk'' modes  may be massless and $\Omega_i^*$ is not IR
finite, requiring itself an IR regularisation.}, it only needs UV
regularisation. However, we cannot remove its $\chi$ dependence 
i.e. set $\chi=0$ in $\Omega_i^*$ of (\ref{QFT3})
before performing the integral. This is because of the 
{\it massless} sector  ($\chi$ dependent)  which  prevents us from doing 
so at this stage, and because the limits $\xi\!\ra \!0$ 
and $\chi\!\ra\! 0$ of $\Omega_i^*$ may not commute.

The $\chi$ independent part of 
the integrand  of eq.(\ref{QFT3}) can be written, using eq.(\ref{poisson1}): 
\begin{eqnarray}\label{cI1}
\!\!\!\!\!\! \cI&\!\!\! \equiv &\!\!\! \sum_{m_{1,2} \in \bZ}' \! 
e^{-\frac{\pi\,t}{T_2 U_2} |U m_1-m_2|^2} 
=\sum_{m_{2} \in \bZ}' \, e^{-\frac{\pi\,t}{T_2 U_2} m_2^2} 
+\sum_{m_{1} \in \bZ}' \sum_{m_2 \in \bZ}\, 
e^{-\frac{\pi\,t}{T_2 U_2}  |U m_1-m_2|^2}\nonumber\\
\nonumber\\
&\!\!\!\!\!\!\!\!\!\!\!\!\!=&\!\!\!\!\!\!\!\!\!\!\!\!
\sum_{m_{2} \in \bZ}' \! e^{-\frac{\pi\,t}{T_2 U_2} m_2^2} 
+\!\bigg[\frac{T_2 U_2}{t}\bigg]^{\frac{1}{2}}\!\!\!
\sum_{m_{1} \in \bZ}' \! e^{-{\pi\,t}\frac{U_2}{T_2} m_1^2}
+\bigg[\frac{T_2 U_2}{t}\bigg]^{\frac{1}{2}}\!\!\!
\sum_{m_{1} \in \bZ}'\sum_{\tilde m_2 \in \bZ}' \!
e^{-{\pi\,t  \frac{U_2}{T_2} m_1^2 -\frac{\pi}{t} T_2 U_2\, \tilde m_2^2
-2 i \pi m_1 \tilde m_2\, U_1}}
\label{cI2}
\end{eqnarray}
A ``prime'' on a single sum stands for the absence of the zero-mode.
$m_2$ was replaced in (\ref{cI2}) by the Poisson re-summed index
$\tilde m_2$. Further,
each of the sums above (multiplied by $exp(-\pi \chi t)$) 
can be integrated over  $t\in [\xi,\infty)$ 
separately. From (\ref{QFT3}), (\ref{cI2}) we  have  
\begin{equation}\label{omegai}
\Omega_i^*=\frac{{\overline {b}}_i}{4\pi}\left(\cJ_1+\cJ_2+\cJ_3\right),
\end{equation}
where 
\begin{eqnarray}
\!\!\!\! \cJ_1&\equiv&\int_{\xi}^{\infty} \frac{dt}{t}
\sum_{m_{2} \in \bZ}' \, e^{-\frac{\pi\,t}{T_2 U_2} m_2^2}\,
e^{- \pi \chi \, t}=-\ln\bigg[4 \pi e^{-\gamma} \, U_2
\frac{T_2}{\xi}\bigg]+2\bigg[\frac{T_2
U_2}{\xi}\bigg]^\frac{1}{2},
\label{dg1}\\
\nonumber\\
\!\!\!\!\!\cJ_2&\equiv&\!\!\!\int_{\xi}^{\infty} \frac{dt}{t}
\bigg[\frac{T_2 U_2}{t}\bigg]^{\frac{1}{2}}
\!\!\!\sum_{m_{1} \in \bZ}' \, e^{-{\pi\,t}\frac{U_2}{T_2} m_1^2}\,
e^{-\pi  \chi \, t}
=\frac{\pi}{3} U_2 +\frac{T_2}{\xi}+\pi T_2 \chi\ln \bigg[4 \pi e^{-\gamma}
\frac{U_2}{T_2} \xi \bigg] -2 \bigg[\frac{T_2
U_2}{\xi}\bigg]^\frac{1}{2}
\label{dg2}\\
\nonumber\\
\!\!\!\!\cJ_3&\equiv&\int_{\xi}^{\infty} \frac{dt}{t}
\bigg[\frac{T_2 U_2}{t}\bigg]^\frac{1}{2}
\sum_{m_{1} \in \bZ}'\sum_{\tilde m_2 \in \bZ}'
e^{-{\pi\,t}\frac{U_2}{T_2} m_1^2-\frac{\pi}{t} T_2 U_2 \tilde m_2^2
-2 i \pi m_1 \tilde m_2 U_1}\, e^{-\pi  \chi \, t}\label{dg3}\\
&=&-\ln \!  \prod_{m_1\geq 1}\left\vert 1-e^{2 i \pi m_1 U}
\right\vert^4.
\label{dg4}
\end{eqnarray}
 $\cJ_1$, $\cJ_2$ and $\cJ_3$ are 
evaluated in detail in the Appendix, see
eqs.(\ref{cj1star}), (\ref{cj2star}) and
(\ref{cj3star}). The above results are only valid in the limit of removing
the UV and IR regulators, $\xi\! \ra \! 0$ and $\chi\!\ra \! 0$. Particular
care must be taken  when the two limits do not ``commute'', in which
case such terms are kept in the final result. The term
$\chi\ln\xi$ in (\ref{dg2}) is  such  an example. 
This  is  the only correction in the 
massive modes contributions   $\cJ_{1}$, $\cJ_2$, $\cJ_3$ compared  
to their value when $\chi=0$  of \cite{Ghilencea:2002ff},
where only the radiative effects of the massive sector alone were 
addressed.

The final result for the total one-loop correction to the tree level 
coupling of eq.(\ref{totalcorrection}) is then (using 
eqs.(\ref{QFT1}) to  (\ref{dg4}))
\begin{equation}\label{finalresult}
\Omega_i^T=\frac{b_i}{4 \pi}
\ln \frac{\Lambda^2}{Q^2}
- \frac{{\overline {b}}_i}{4 \pi} 
\ln\Big[4 \pi e^{-\gamma} \, e^{-{T_2^*}} \, {T_2^*} \, U_2
\, \vert \eta(U)\vert^4\Big]
+\frac{{\overline {b}}_i}{4 \pi}Q^2 R_1 R_2 \sin\theta \, e^{-\gamma}
\ln \bigg[4 \pi e^{-\gamma} \frac{U_2}{T_2^*}\bigg]
\end{equation}
We denoted
\begin{equation}\label{cutoff}
T_2^*\equiv\left.\, \frac{T_2}{\xi}\, \right\vert_{\xi\ra 0}
= \Lambda^2  R_1 R_2 \sin\theta, \qquad \quad \textrm{and}
\qquad 
\left.\Lambda^2\equiv\frac{\mu^2}{\xi}\right\vert_{\xi\ra 0}, 
\qquad Q^2\equiv \pi e^{\gamma}
\, \mu^2 \chi\, \bigg\vert_{\chi\ra 0}
\end{equation}
$Q$ is  associated with a low-energy scale and $\Lambda$ is
the UV scale. Eq.(\ref{cutoff}) also clarifies the link between the
UV/IR regulators and their associated mass scales.

The main result (\ref{finalresult}) is valid in the limit
of removing the regulators when higher order corrections in
these are vanishing (see  Appendix D eqs.(\ref{rmreg1}), 
(\ref{rmreg2}), (\ref{rmreg3})).
For the  mass scales $Q$ and $\Lambda$ of the model, this means that
\begin{eqnarray}\label{conditions1}
Q \ll \frac{1}{R_2 \sin\theta} \leq  \frac{1}{R_1}\ll \Lambda
,\,\,\,\,\, (U_2\geq 1)\qquad \textrm{or}\qquad 
Q \ll \frac{1}{R_1} \leq  \frac{1}{R_2\sin\theta }\ll
\bigg[\frac{\Lambda}{R_1}\bigg]^{1/2}\!\!\!\! \ll \Lambda
,\,\,\,\,\, (U_2 < 1)
 \end{eqnarray}
from which  one can also find $T_2^*\equiv
\Lambda^2 R_1 R_2 \sin\theta\gg 1$. $Q$ is the lowest mass scale of
the theory.

Eq.(\ref{finalresult}) shows an interesting result. The first term
is just the massless modes logarithmic contribution from 
the UV cut-off  ($\Lambda$) to the low energy scale ($Q$) where such
states  decouple. 
It corresponds to the second term in the r.h.s. of (\ref{gauge}).
 The second term in (\ref{finalresult})
is the contribution (IR finite, UV regularised) of 
{\it massive} Kaluza-Klein modes alone mentioned in the introduction, 
eq.(\ref{t2form}) in the string case and eq.(\ref{result}) in 
the EFT approach (with $C_{reg.}=4\pi \exp(-\gamma)$).
Its relationship  to the result of
the heterotic string case \cite{Dixon:1990pc} 
was extensively discussed in \cite{Ghilencea:2002ff}.

The last  term in (\ref{finalresult}) 
is a new correction logarithmic in the UV scale $\Lambda$ 
and is not present in the limit $\alpha'\!\ra\! 0$ of the
string calculation eq.(\ref{t2form}) or previous effective field
theory calculations \cite{Ghilencea:2002ff}, 
\cite{Taylor:1988vt}. It is due to the term $\pi T_2
\chi\ln \xi$  of eq.(\ref{dg2}) which is {\it relating/mixing the IR  and UV 
scales through their associated regulators $\chi$ and $\xi$
respectively}. 
This term arises from the sector of  massive Kaluza-Klein modes in 
the presence of the IR regulator $\chi$. 
The  infrared dependent part of the massive (``bulk'')  modes 
 is controlling (part of) the UV behaviour of the gauge couplings. 
The effect is related to  the infinite number of momentum states that 
we summed over, see also its origin in  the first integral in 
(\ref{j2star}), and in  (\ref{cj2starxi0}), (\ref{cj2starxi}) of 
Appendix B. These equations also show the effect is present in
models with {\it two}  additional compact  dimensions
(more generally, for an even number of these).  
The effect arises from  Poisson re-summed
(0,0) levels\footnote{Note that these levels are different from 
the ``original'' (0,0) Kaluza-Klein  modes.} with respect to both 
dimensions compactified on  the two-torus. These are the levels which 
also bring the leading power-like correction to the couplings.

The last term in (\ref{finalresult}) also shows that even though the 
massive Kaluza-Klein modes may have a large  mass, of the order the
 compactification scale, they can still bring a contribution
proportional to the (much) lower  scale $Q$ (where they are actually
decoupled!). The logarithm can be large since $U_2\ll T_2^*$ 
(equivalently $1/R_1\ll \Lambda$). However the coefficient in 
front is small $Q^2 R_1 R_2 \sin\theta\ll1$ for  our result to 
hold, see (\ref{conditions1}), and vanishes when the two cycles 
of the torus collapse onto each other $\theta\ll 1$, (one extra 
dimension case). The new correction  implies the existence in some 
compactified effective field theories of a connection between the 
IR and UV sectors. This  effect is due to the presence of the
IR regulator (required by  the massless sector) combined  with the 
UV divergence due to infinitely many massive states.  
It seems that the overall  effect of
the {\it infinite} Kaluza-Klein tower on the gauge couplings
 cannot be split into massive and massless contributions only,
as done in eq.(\ref{gauge}), and an additional  term is present, 
the last term in eq.(\ref{finalresult}).
This term  is ultimately  a combined effect of these two mass 
sectors, through infrared effects.

There remains the difficult question of regularisation independence. 
The UV  regularisation employed here is supported by previous works 
\cite{Ghilencea:2002ff} which  match the string result
\cite{Dixon:1990pc} for the corrections due to  massive modes 
alone (no massless state present). 
The IR regularisation using $exp(-\chi t), \chi\!\ra\! 0$ 
is similar to the IR regularisation
used in some string theory approaches  \cite{Kiritsis:1994yv}. 
The discussion in the next section also gives support to the 
regularisation choice independence of the result obtained.

The new correction found may have interesting implications. 
At the phenomenological level an example would be its effects on 
the exact value of the unification scale in the class of models 
addressed here, and the matching of the MSSM  unification scale vs. 
the  heterotic string scale.

%%%%%%%%%%%%%%%%%%%%%%%%%%%%%%%%%%%%%%%%%%%%%%%%%%%%%%%%%%%%%%%%
%\vspace{0.8cm}
\section*{The limit $\alpha'\ra 0$ of the string.}

%%%%%%%%%%%%%%%%%%%%%%%%%%%%%%%%%%%%%%%%%%%%%%%%%%%%%%%%%%%%%%%%

The presence of the new correction  found at the EFT level may have
additional implications, in particular on the link between 
this  approach  and string theory. At the (heterotic) string level 
the role of the UV scale $\Lambda$
(or equivalently $\xi\ra 0$) is replaced  by the string scale $M_s$ or
equivalently  $\alpha'\propto 1/M_s^2$. In the limit $\alpha'\ra 0$ one would 
expect that the string  result and the effective field theory result 
(\ref{finalresult}) would match.
This does actually happen for the effects of the {\it massive} modes
alone,  $\Omega_i^H$ and $\Omega_i^{EFT}$ while ignoring the massless modes
role in the theory,  as discussed  in the introduction, eqs.(\ref{gauge}), 
(\ref{t2form}), (\ref{result}) and in reference \cite{Ghilencea:2002ff}.

The difference between the (limit $\alpha'\!\ra \!0$ of the) 
string results and those of the EFT approach
appears when considering the role of the {\it massless} states.
At the EFT level they introduced the IR regulator whose presence 
in the massive sector brought in the last term in (\ref{finalresult}).
This term  is not recovered from 
available string results in the limit $\alpha'\!\ra\! 0$, 
(see eq.(\ref{gauge}) and (\ref{t2form})). 
To understand why this is so, it is important to mention that the string 
evaluation of the corrections to the  gauge couplings requires an infrared 
regularisation \cite{Dixon:1990pc}. This is  because
the contribution of the  degenerate orbits of the modular group 
SL(2,Z) contributing to the gauge couplings (see Appendix of 
\cite{Dixon:1990pc}) is IR divergent. In the dimensional
regularisation version \cite{Foerger:1998kw}
of the (IR divergent) string calculation, the {\it total} string correction 
to the ``bare'' gauge couplings\footnote{In this work we
always refer to the {\it gauge dependent} part of the string
correction to the gauge couplings. The universal part (gauge independent) 
 is ``absorbed''  into the definition
of the ``bare'' or string coupling $\alpha_o$ of eq.(\ref{gauge}) to
ensure this is invariant (as it should) under the symmetries of the string,
$SL(2,Z)_T\times SL(2,Z)_U\times
Z_2^{T\leftrightarrow U}$ \cite{Nilles:1997vk}.}
is  evaluated to (the beta functions are not shown explicitly)
\begin{equation}\label{string_h}
\Omega^T_{string} \propto 
\lim_{\epsilon\ra 0} \int_{\Gamma} \frac{d^2
 \tau}{\tau_2^{1+\epsilon}} Z\propto
\bigg[ \frac{1}{\epsilon} +
\Omega^H + \epsilon \ln (T_2/\alpha')
+\cO(\epsilon)\bigg]_{\epsilon\ra 0}
\end{equation}
$\epsilon$ is the infrared regulator in string case ($\epsilon\ra 0$),
$Z$ the string partition function including the 
{\it  massless} (``zero'') modes contribution, and with the $\alpha'$
dependence of $T_2$ as defined in string case, shown explicitly.  
$1/\epsilon$  accounts
for the massless modes infrared divergence, which is logarithmic  in scale, 
and included in  eq.(\ref{gauge}). Finally $\Omega^H $ is just
$\Omega_i^H/\overline b_i$  with $\Omega_i^H$ given  in eq.(\ref{t2form}), 
and stands for the {\it massive} modes contribution.
Additional terms like  $\epsilon\ln(T_2/\alpha')$
with $\epsilon\ra 0$  do appear in (\ref{string_h}) during the 
IR regularisation of the  string (see details in Appendix A of 
\cite{Foerger:1998kw}) and are similar in structure to  
$\chi\ln \xi$ of eq.(\ref{dg2}) leading to the last term in 
(\ref{finalresult}). Other terms 
relevant for the field theory limit $\alpha'\!\ra \!0$, may be 
present in eq.(\ref{string_h}) e.g.  $\epsilon (T_2/\alpha')$ 
or may   have a more complicated structure like 
$\epsilon \times g(T)$ with $g(T)$ being  some $SL(2,Z)_T$ invariant
function and $\epsilon\!\ra \! 0$.

In the limit of removing the IR string regulator $\epsilon\ra 0$,
the term  $\epsilon\ln(T_2/\alpha')$ of eq.(\ref{string_h})
vanishes and is not present in the final string correction to the
gauge couplings given by eqs.(\ref{gauge}), (\ref{t2form}).
This  is  true  only if $\alpha'$ is  {\it non-zero}
as assumed by  string calculations.   
However, in the  limit  $\alpha'\ra 0$ supposed to give the 
field theory regime,  such term 
  can bring corrections to the gauge 
%%%%%%%%%%%%%%%%%%
couplings\footnote{There are other IR  string regularisations, considered 
to be modular invariant \cite{Kiritsis:1994yv}
which introduce  a mass gap parameter $\mu^2/M_s^2\ll 1$ as IR 
regulator. The correspondent of this regulator  in our  EFT approach 
is $\chi$. In this IR string regularisation it would be useful to know  
if a  calculation of the {\it massive} modes effects
gives [in addition to $\Omega_i^H$ of (\ref{t2form})],  correction
terms $\cO(\mu^2/M_s^2)$ of structure 
$\mu^2/M_s^2 (T_2/\alpha')\ln(T_2/\alpha')\propto \mu^2 R_1 R_2 \sin
\theta \ln(T_2/\alpha')$. Such $\cO(\mu^2/M_s^2)$ terms were not 
computed in this regularisation scheme.} 
%%%%%%%%%%%%%%%%%%%
as it was the case in the EFT analysis, and should be added to the
final result.
Therefore the  limits $\alpha'\!\ra\! 0$ and $\epsilon\!\ra\! 0$ do
not commute (similar to the EFT case where $\xi\!\ra\! 0$ and
$\chi\!\ra\!0$ limits  on $\chi\!\ln \xi$  did not commute). This 
also requires an analysis at the string level,
to explain if any  string symmetry can still prescribe (upon
compactification) the order to take such limits in eq.(\ref{string_h}).
In a sense this  tells us that to understand the UV behaviour of 
an effective field theory  model  from the limit $\alpha'\ra 0$ of 
its  string embedding, one may have to address the IR behaviour 
of the string!

We think that one can  recover in the string case the additional 
correction  found in the EFT approach, the last term in 
eq.(\ref{finalresult}), using the regularisation approach of
\cite{Kiritsis:1994yv}.  At string level this 
correction may  have a  more general form, invariant under string
symmetries left after compactification. The limit $\alpha'\!\ra \!0$
on such additional string correction should recover the last term in 
eq.(\ref{finalresult}) computed in the EFT approach.

The above discussion explains why there is no 
string counterpart  available for the last term in
(\ref{finalresult}), 
even though  terms of  structure similar to it can arise in the 
string case. This supports  our expectation that the last term in
(\ref{finalresult})  is not a result of  the regularisation choice, 
but  a generic presence in the EFT approach.
 This  gives an example where   one cannot recover from 
the $\alpha'\!\ra\! 0$ limit of available string results
the UV behaviour as obtained on pure EFT grounds.
The effect may be associated with the non-renormalisability
of the EFT approach and the {\it infinite} number of momentum states. 
The extra term found in the EFT case raises questions on the exact 
matching procedure of the $\alpha'\!\ra\! 0$ limit of some
(infrared  regularised) string calculations with effective field 
theory results.

\vspace{0.2cm}
\section*{Conclusions}
The effects of two extra dimensions compactified on a
two-torus were studied in an effective field theory approach
by summing  one-loop radiative corrections 
to the gauge couplings, due to massless and massive
Kaluza-Klein modes. First, if one considered the
one-loop correction to the gauge couplings  of the massive modes alone  
(i.e. if no massless states existed in the theory) one would only need 
an UV regularisation.  The result obtained would be equal  to that of 
the heterotic string in the limit $\alpha'\!\ra\! 0$, as previous 
works showed. However when including in the field theory approach
the effects of the  massless states of 
the theory  an IR regularisation is required in addition to the 
UV regularisation. A consistent treatment of these regularisations
brings an additional correction   to the gauge couplings,
which is logarithmic in the UV scale.  This (one-loop) correction 
originates from a  {\it UV-IR mixing term} 
$\chi\ln\xi$ (with $\chi\!\ra\! 0$, $\xi\!\ra \!0$)  
{\it between the UV and IR sectors} through their associated
regulators $\xi$ and $\chi$ respectively.
Consequently, one obtains the surprising result that these sectors 
do not decouple from each other and this is due to the presence of 
{\it infinitely}  many Kaluza-Klein states. 
This also shows that the  UV behaviour of the couplings is sensitive  
to IR regulator   effects associated with  massive (``bulk'') modes. 
The effect is present for two compact dimensions.

Regarding  the link with string theory, we discussed why the new 
field theory  correction  is not recovered from  known heterotic 
string results in  the limit $\alpha'\!\!\ra\!  0$.
While evaluating this limit, one should  note that
string calculations for radiative corrections (to gauge couplings)
usually require an IR regularisation. In the limit of 
removing the IR string regulator $\epsilon\! \ra \!0$, one
discards those $\alpha'$  dependent terms which  are
multiplied by the regulator, for example $ \epsilon\ln\alpha'$,
($\alpha'$ non-zero).  Such terms are relevant for the gauge
couplings, at least in the limit 
$\alpha'\ra 0$ of the string. This is because such $\alpha'$ dependent 
terms considered on their own, without the $\epsilon$ dependence
 are  divergent  in the limit $\alpha'\!\ra\! 0$ supposed to give 
the field theory regime. 
The discrepancy between the EFT and  the limit $\alpha'\!\ra\! 0$ of the 
string is ultimately due to the fact that the
limits $\alpha'\!\ra\! 0$  and $\epsilon\!\ra\! 0$  of the string do not
commute. This issue and its implications deserve further investigation.

The effective field theory result applies to  models with 
``large'' extra dimensions and  is  valid without reference to  
string theory. The additional radiative correction to the gauge couplings
that was found  shows a connection via infrared regulator effects  
between the massless and massive sectors (of infinitely many states) 
of a compactified theory. 
The example discussed  tells us  that the limit $\alpha'\!\ra\! 0$ on
the results of the (IR regularised) string  as well as its matching
with corresponding effective field theory results is
more subtle than initially thought, if massless states are present 
in the theory.

\vspace{1.3cm}
 \noindent
 {\bf Acknowledgements:} 

\noindent
It is a pleasure  to thank H.P. Nilles and F. Quevedo for 
helpful discussions and observations on this work. The author 
thanks  S.~Stieberger for his many helpful suggestions and comments 
on the manuscript. He also thanks
 S. Groot-Nibbelink for discussions on related topics. 
The author acknowledges the  financial support from PPARC (UK).

%%%%%%%%%%%%%%% Appendix %%%%%%%%%%%%%%%%%%%%%%%%%%%%%%%%%%%%%%%%%

\def\baselinestretch{1.25}

%%%%%%%%%%%%%%%%%%%%%%%%%%%%%%%%%%%%%%%%%%%%%%%%%%%%%%%%%%%%%%%%%%

%%%%%%%%%%%%%%%%%%%%%%%%%%%%%%%%%%%%%%%%%%%%%%%%%%%%%%%%%%%%%%%%%%
%%%  J1
%%%%%%%%%%%%%%%%%%%%%%%%%%%%%%%%%%%%%%%%%%%%%%%%%%%%%%%%%%%%%%%%%%

\newpage
\section*{Appendix}
\appendix
\setcounter{equation}{0}
\def\theequation{A-\arabic{equation}} 
\subsection*{A. Kaluza-Klein Integrals:  $\cJ_1$}
General integrals present in Kaluza-Klein models are evaluated below 
(see also Appendix C in \cite{Ghilencea:2001bv}).

\noindent
$\bullet$ We evaluate the integral: ($\beta>0$, $y>0$)
\begin{equation}\label{cj1star}
\cJ_1^*(\beta, y)\equiv 
\int_{\xi}^{\infty}\frac{dt}{t} \, \sum_{m_2}' e^{-\pi\, \beta\, t\,
(m_2^2+y)}
\end{equation}
in the limit $\xi\ra 0$ and $y\ra 0$. \\
\noindent
For $\beta=1/(T_2 U_2)$, $y=\chi/\beta$
one recovers the integral $\cJ_1$ of eq.(\ref{dg1}) in
the text. \\

\noindent
$\cJ_1^*$ does not have divergences in $y\!\ra\! 0$, 
(see eq.(A-1) in \cite{Ghilencea:2002ff} where $y\!=\!0$ 
case is evaluated) thus terms $\xi/y$ or $\xi\ln y$ and alike are 
not present or ``missed'' when in the  second integral in
(\ref{sl2}) we set $\xi=0$. For the error introduced by doing so,
 see  eq.(\ref{delt1}). We have ($\xi\ra 0$)
\begin{eqnarray}
\!\!\!\!\!\!\!\!\!\!\!\!\cJ_1^*&=&
 \int_{\xi}^{\infty} \frac{dt}{t}\bigg[-1  +\frac{1}{\sqrt{\beta t}}
+\frac{1}{\sqrt{\beta t}}\sum_{\tilde m_2}'e^{-\frac{\pi}{\beta t}
\tilde m_2^2}\bigg] e^{-\pi \beta\, y\, t}\label{sl1}\\
&=& \int_{\xi}^{\infty} \frac{dt}{t}\bigg[-1  +\frac{1}{\sqrt{\beta
t}}\bigg] e^{-\pi \beta \, y t}+\frac{1}{\sqrt \beta}\sum_{\tilde m_2}' \int_{0}^{\infty}
\frac{dt}{t^{3/2}} \, e^{-\frac{\pi}{\beta t} 
\tilde m_2^2-\pi \beta t y}\label{sl2}\\
&=& -\Gamma[0,\pi\beta \xi y]+\frac{2}{\sqrt{\beta \xi}} e^{-\pi \beta \xi y}
-2\pi \sqrt{y} \bigg[1-\textrm{Erf}[ \sqrt{\pi \beta \xi y}]\bigg]
-2 \ln\bigg[1-e^{-2\pi \sqrt  y}\bigg]\label{j1star}
\end{eqnarray}
In (\ref{sl1}) we used 
\begin{equation}\label{poisson1}
\sum_{n \in \bZ} e^{-\pi A (n+\sigma)^2} = \frac{1}{\sqrt A}
\sum_{\tilde n\in \bZ} e^{-\pi A^{-1} {\tilde n}^2  + 2 i \pi  
{\tilde n}  \sigma} 
\end{equation}
We also used 
\begin{equation}
\textrm{Erf}[x]\equiv \frac{2}{\sqrt{\pi}}\int_{0}^{x}dt\, e^{-t^2};
\quad \textrm{Erf}[x]=
\frac{2 x}{\sqrt\pi}-\frac{2
x^3}{3\sqrt\pi}+\cO(x^5),\quad\textrm{if  } (x\ll 1)
\end{equation}
and  \cite{gr}
\begin{equation}\label{bessel1}
\int_{0}^{\infty} dx\, x^{\nu-1} e^{- b x^p- a
x^{-p}}=\frac{2}{p}\, \bigg[\frac{a}{b}
\bigg]^{\frac{\nu}{2 p}} K_{\frac{\nu}{p}}(2 \sqrt{a \, b}),\quad Re
(a), (b)>0,\qquad K_{-\frac{1}{2}}(z)=\sqrt{\frac{\pi}{2 z}} \, e^{-z}
\end{equation}
From (\ref{j1star}) we find the leading behaviour
\begin{equation}\label{finalj1star}
\cJ_1^*(\beta, y)=-
\ln \bigg[ 4 \pi e^{-\gamma} \frac{1}{\beta \xi}\bigg]+\frac{2}{\sqrt{\beta
\xi}},\qquad \xi\ra 0,\, y\ra 0.
\end{equation}
(\ref{finalj1star}) is a good approximation to
(\ref{j1star}) if $\xi$ and $y$  respect additional conditions, converted
into  constraints on the mass scales, see Appendix D
eq.(\ref{rmreg2}). The two limits  $\xi\!\ra\! 0$  and 
$y\!\ra\! 0$ on (\ref{cj1star}) commute (see also eq.(A-1) of the 
Appendix of \cite{Ghilencea:2002ff}). 
Eqs.(\ref{cj1star}), (\ref{finalj1star}) 
give $\cJ_1$ used in the text.

%%%%%%%%%%%%%%%%%%%%%%%%%%%%%%%%%%%%%%%%%%%%%%%%%%%%%%%%%%%
%%% J2
%%%%%%%%%%%%%%%%%%%%%%%%%%%%%%%%%%%%%%%%%%%%%%%%%%%%%%%%%%%

\setcounter{equation}{0}
\def\theequation{B-\arabic{equation}} 
\subsection*{B. Kaluza-Klein Integrals:  $\cJ_2$}
\noindent
$\bullet$ We  evaluate the integral ($\beta>0$, $y>0$)
\begin{equation}\label{cj2star}
\cJ_2^*(\beta, y)\equiv \int_{\xi}^{\infty}\frac{dt}{t^{3/2}}\sum_{m_1}'
e^{-\pi \beta \, t\,  (m_1^2+y)}
\end{equation}
in the limits $\xi\ra 0$, $y\ra 0$. \\
\noindent
For $\beta=U_2/T_2$ and $y=\chi/\beta$
% $y=Y\, T_2/U_2$ 
one recovers $\cJ_2$ of eq.(\ref{dg2}) up to an overall factor,
 $\cJ_2=\sqrt{T_2 U_2}\cJ_2^*$. \\

\noindent
 $\cJ_2^*$ does not have divergences in $y\!\ra \! 0$, 
(see eq.(A-12) of \cite{Ghilencea:2002ff} where $y\!=\! 0$ case is
evaluated); thus 
terms divergent in $y$ for example
 $\xi/y$ or $\xi\ln y$ are not present or
 ``missed'' when in the second integral in (\ref{j2s2}) we  set
$\xi=0$. For the error introduced by doing so see eq.(\ref{delt2}).
We have ($\xi\ra 0$):
\begin{eqnarray}
\!\!\!\!\!\!\!\!\!\!\!\!\!
\cJ_2^*&=& \int_{\xi}^{1} \frac{dt}{t^{3/2}}
\bigg[-1+\frac{1}{\sqrt{\beta t}} \sum_{\tilde m_1} e^{-\pi
\frac{\tilde m_1^2}{\beta t}}\bigg]\, e^{-\pi \beta t y}
+\int_{1}^{\infty} \frac{dt}{t^{3/2}}
\sum_{m_1}'e^{-\pi \,\beta\, t\, ( m_1^2+y)}\label{j2s1}\\
&=& \!\! \int_{\xi}^{1} \frac{dt}{t^{3/2}}
\bigg[\frac{1}{\sqrt{\beta t}}-1\bigg]
\, e^{-\pi \beta\, y\, t} +\int_{1}^{1/\xi} \frac{dt}{\sqrt \beta}
 \sum_{\tilde m_1}' e^{-\pi t \frac{\tilde m_1^2}{\beta}} 
e^{-\pi \beta\, y/t}+\int_{1}^{\infty}\frac{dt}{t^{3/2}}
\sum_{m_1}' e^{-\pi \beta t (m_1^2+y)}
\label{j2s2}\\
&=&\!\!\!\!
\int_{\xi}^{1} \frac{dt}{t^{3/2}}\bigg[
\frac{1}{\sqrt{\beta t}}-1 \bigg]\, e^{-\pi \beta\, y\, t}
+\! \int_{1}^{\infty}\! dt 
\bigg[\frac{1}{\sqrt \beta}\sum_{\tilde m_1}' e^{-\pi t \frac{\tilde
m_1^2}{\beta}} e^{-\pi \beta\, y/t}+\frac{1}{t^{3/2}}
\sum_{m_1}' e^{-\pi \beta t (m_1^2+y)}\bigg]%\nonumber\\
\label{j2star}
\end{eqnarray}
We denote the first integral in (\ref{j2star}) by $\cJ_{2, \xi}^*$ 
while the second integral (finite) can be written as a
well defined limit $\lim_{\epsilon\ra 0} \cJ_{2, \epsilon}$ 
with  $\cJ_{2,\epsilon}$ defined/computed in (\ref{cj2epsilon}). 
For  $\cJ_{2,\xi}^*$ we have
\begin{eqnarray}
\cJ_{2,\xi}^*&\equiv & \int_{\xi}^{1} \frac{dt}{t^{3/2}}\bigg[
\frac{1}{\sqrt{\beta t}}-1 \bigg]\, e^{-\pi \beta\, y\, t}\nonumber\\
&=& (\pi \beta y)^{\frac{1}{2}}\bigg[
\Gamma[-1/2,\pi \beta  y]-\Gamma[-1/2, \pi\beta  y \xi]\bigg]
-\frac{e^{-\pi \beta y}}{\beta^{1/2}}+\frac{e^{-\pi\beta \xi y}}{\xi
\beta^{1/2}}\nonumber\\
&-& \pi y  \beta^{\frac{1}{2}}\bigg[\textrm{Ei}[-\pi\beta y]-\textrm{Ei}[-\pi \beta y \xi]\bigg]\label{cj2starxi0}
\end{eqnarray}
Using that \cite{gr}
\begin{equation}\label{expint}
\textrm{Ei}[-z]=\gamma+\ln z+\sum_{k\geq 1}\frac{(-z)^k}{k!\,
k},\, \textrm{  for } z>0
\end{equation}
\begin{equation}\label{gammas}
\Gamma[\alpha]-\Gamma[\alpha,x]
=x^{\alpha} \sum_{n\geq 0} \frac{(-1)^n x^{n}}{n! \, (\alpha+n)},
\qquad\alpha\not=0,-1,-2,\cdots
\end{equation}
one can expand (\ref{cj2starxi0}) for $\xi$ and $y$  small enough,  
to find the leading behaviour 
\begin{equation}\label{cj2starxi}
\cJ_{2,\xi}^*=
 \beta^{\frac{1}{2}} \bigg[\frac{1}{\xi\beta}-\frac{2}{\sqrt{\xi \beta}}
+\pi  \, y \ln\xi-\frac{1}{\beta}
+\frac{2}{\sqrt\beta}\bigg],\qquad \xi\ra 0, \, y\ra 0.
\end{equation}
For the term $y\ln\xi$ the limits $y\ra 0$ and $\xi\ra 0$ 
do not commute and it is for this reason that this term is kept in
(\ref{cj2starxi}). This is the only such case. 
This term is a mixing of IR/UV contributions, 
with implications discussed in the text. Its origin is traced to 
$\int_{\xi}^{1}dt/t^2 exp(-\pi \beta y t)$ of
eq.(\ref{j2star}) arising from ``Poisson re-summed (0,0)''
Kaluza-Klein modes with respect to {\it both} dimensions.

In the step from (\ref{j2star}) to (\ref{cj2starxi}) 
additional corrections (convergent series)
are discarded. These can be neglected if $\xi$ and $y$ 
respect additional conditions  converted into bounds on 
the  mass scales of the theory in Appendix~D eq.(\ref{rmreg3}).\\

\vspace{0.3cm}
\noindent
We now compute the last integral in eq.(\ref{j2star}) which is equal to 
the finite limit $\lim_{\epsilon \ra 0} \cJ_{2, \epsilon}$ with
\begin{eqnarray}
\!\!\!\!\!\!\!\!\!\!\!\!\!\! \!\!\!\! 
\cJ_{2,\epsilon}^* & \equiv& \int_{1}^{\infty} dt
\bigg[ \frac{t^\epsilon}{\sqrt \beta}\sum_{\tilde m_1}' e^{-\pi t \frac{\tilde
m_1^2}{\beta}} e^{-\pi \beta\, y/t}+\frac{1}{t^{3/2+\epsilon}}
\sum_{m_1}' e^{-\pi \beta t (m_1^2+y)}\bigg]\nonumber\\
&=&\frac{1}{\sqrt \beta} \int_{0}^{1} \frac{dt}{t^{2+\epsilon}} 
\sum_{\tilde m_1}' e^{-\pi \frac{\tilde
m_1^2}{\beta t}} e^{-\pi \beta\, y \, t}+\int_{1}^{\infty}
\frac{dt}{t^{3/2+\epsilon}}
\sum_{m_1}' e^{-\pi \beta t \,(m_1^2+y)}\nonumber\\
&=&\!\! \frac{1}{\sqrt \beta} \int_{0}^{1}\!\! \frac{dt}{t^{2+\epsilon}} 
\bigg[\!-1+(\beta \, t)^{\frac{1}{2}}\bigg] e^{-\pi \beta \, y \, t}
+\int_{0}^{\infty}\!\!\!
\frac{dt}{t^{3/2+\epsilon}}
\sum_{m_1}' e^{-\pi \beta t \,(m_1^2+y)}\nonumber\\
&\equiv&  \cM_1+\cM_2\label{cj2epsilon}
\end{eqnarray}
where $\cM_{1,2}$ denote the first/second integral respectively. 
The second integral is similar to a DR regularised version
of the initial integral  (\ref{cj2star}), for  details
see also Appendix B, C of \cite{Ghilencea:2002ff}.

To evaluate the first integral  ($\cM_1$) of (\ref{cj2epsilon})  we have
\begin{eqnarray}
\cM_1&\equiv&
\frac{1}{\sqrt \beta} \int_{0}^{1} \frac{dt}{t^{2+\epsilon}} 
\bigg[-1+(\beta \, t)^{\frac{1}{2}}\bigg] e^{-\pi \beta \, y \, t}\nonumber\\
%\nonumber\\
&=& (\pi \beta y)^{\epsilon+\frac{1}{2}}
\bigg\{ (\pi\,y)^{\frac{1}{2}}  \bigg[\Gamma[-1-\epsilon,\pi\beta
y]-\Gamma[-1-\epsilon]\bigg]+
\Gamma[-1/2-\epsilon]-\Gamma[-1/2-\epsilon,\pi \, \beta  \,y]\bigg\}\nonumber\\
%\nonumber\\
% &=&\frac{1}{\sqrt\beta}\bigg[\frac{1}{1+\epsilon}-\frac{\pi\beta
% y}{\epsilon}-\sum_{n=2}^{\infty} \frac{(-\pi \beta y)^n}{n! \, 
% (n-1-\epsilon)}\bigg]
% +\frac{-1}{1/2+\epsilon}+\sum_{n=1}^{\infty}\frac{(-\pi \beta y)^{n}
% }{n! \, (n-1/2-\epsilon)}\nonumber\\
% \nonumber\\
&=&\beta^{\frac{1}{2}} \bigg[\frac{1}{\beta}-\frac{2}{\sqrt\beta}-\frac{\pi\,
y}{\epsilon}\bigg]- \frac{1}{\sqrt\beta}
\sum_{n=2}^{\infty} \frac{(-\pi \beta y)^n}{n! \, (n-1)}
+\sum_{n=1}^{\infty}\frac{(-\pi \beta y)^{n}
}{n! \, (n-1/2)}+\cO(\epsilon)\label{eq10}
\end{eqnarray}
We used eq.(\ref{gammas}) to isolate the
only ``mixing'' term $\pi \, y/\epsilon$ from the remaining 
(convergent) series.

To evaluate the second integral ($\cM_2$) of (\ref{cj2epsilon}) we have:
\begin{eqnarray}
\cM_2& \equiv &\int_{0}^{\infty}
\frac{dt}{t^{3/2+\epsilon}}
\sum_{m_1}' e^{-\pi \beta\, t \,(m_1^2+y)}
=2 \, \Gamma[-1/2-\epsilon] (\pi \beta)^{\frac{1}{2}+\epsilon}
\sum_{m_1>0}  \frac{1}{(m_1^2+y)^{-1/2-\epsilon}}\nonumber\\
\nonumber\\
&=& 2 \,  (\pi \beta)^{\frac{1}{2}+\epsilon}
\sum_{m=0}^{\infty} \frac{(-y)^m}{m!} \,\Gamma[m-1/2 -\epsilon] 
\, \zeta(2m-1-2\epsilon)\label{limit}
\end{eqnarray}

\noindent
In the last step we used the convergent expansion \cite{elizalde} ($0<y<1$)
\begin{equation}
\sum_{n>0} \frac{1}{(n^2+y)^s}=\sum_{m=0}^{\infty}
\frac{\Gamma[m+s]}{\Gamma[s] \, m!}\, ({-y})^m \,\zeta(2
s+2 m)
\end{equation}
In (\ref{limit})  the terms with $m\not=1$ give finite
contributions for $\epsilon\ra 0$ and   $0<y<1$ (in these we can 
set  $\epsilon=0$).  However,  the term with $m=1$ brings 
$\zeta(1-2\epsilon)$, singular in $\epsilon\ra 0$. Expanding in 
$\epsilon$ the factors of this term leads to 
\begin{eqnarray}
\zeta(1-2 \epsilon)&=&-\frac{1}{2 \epsilon}+\gamma+\cO(\epsilon)\nonumber\\
\Gamma[1/2-\epsilon]&=&\pi^\frac{1}{2}(1-\epsilon\,
\psi(1/2)+\cO(\epsilon^2)),\qquad \psi(x)\equiv d(\ln\Gamma[x])/dx\nonumber \\
(\pi \beta)^{\epsilon}&=& 1+\epsilon \ln(\pi\beta)+\cO(\epsilon^2)
\end{eqnarray}
Altogether we find for $\cM_2$
\begin{eqnarray}\label{M2}
\cM_2&=&\beta^{\frac{1}{2}}\bigg[\frac{\pi}{3}  +
\frac{\pi\, y}{\epsilon}\bigg]
+\pi \beta^{\frac{1}{2}} \, y \,\ln\left(4 \pi e^{-\gamma}
\beta\right)\nonumber\\
&+&2(\pi\beta)^{\frac{1}{2}}\,\sum_{m=2}^{\infty} 
\frac{(-y)^m}{m!} \,\Gamma[m-1/2] 
\, \zeta(2m-1)+\cO(\epsilon)
\end{eqnarray}
We find again a term $\pi y/\epsilon$, 
cancelled in $\cM_1+\cM_2$. This is a DR ``equivalent'' of
$y\ln\xi$ of (\ref{cj2starxi}).

Adding $\cM_1$ of (\ref{eq10}) and $\cM_2$ of (\ref{M2}) 
we find the finite value of $\lim_{\epsilon\ra 0}\cJ_{2,\epsilon}$ 
of  eq.(\ref{cj2epsilon})
\begin{eqnarray}\label{finalcj2starep}
\lim_{\epsilon\ra 0}\cJ_{2,\epsilon}^*&=&
\cM_1+\cM_2\nonumber\\
&=&\beta^{\frac{1}{2}}
\bigg[\frac{1}{\beta}-\frac{2}{\sqrt\beta}+\frac{\pi}{3}\bigg]- 
\frac{1}{\sqrt\beta}
\sum_{n=2}^{\infty} \frac{(-\pi \beta y)^n}{n! \, (n-1)}
+\sum_{n=1}^{\infty}\frac{(-\pi \beta y)^{n}
}{n! \, (n-1/2)}\nonumber\\
&+&\!\!\!\pi \beta^{\frac{1}{2}} \,y\, \ln\left(4 \pi e^{-\gamma}
\beta\right)+2(\pi\beta)^{\frac{1}{2}}\!\!\sum_{m=2}^{\infty} \frac{(-y)^m}{m!}
 \,\Gamma[m-1/2] \, \zeta(2m-1),\,\, 0\!<\!y\!<\!1
\end{eqnarray}
Eqs.(\ref{cj2starxi0}) and  (\ref{finalcj2starep}) give the 
result for $\cJ_2^{*}$ when   $0<y< 1$.

\noindent
The leading term  of  eq.(\ref{finalcj2starep}) is
\begin{eqnarray}\label{B16}
\lim_{\epsilon\ra 0}\cJ_{2,\epsilon}^*=
\beta^{\frac{1}{2}}\bigg[\frac{1}{\beta}-
\frac{2}{\sqrt\beta}+\frac{\pi}{3}+\pi y \ln\left(4 \pi e^{-\gamma}
\beta\right)   \bigg],\qquad  y\ra 0
\end{eqnarray}
The leading term in  $\cJ_2^*$  is (using 
(\ref{cj2starxi}), (\ref{B16})) 
\begin{equation}\label{j2starxi}
\cJ_{2}^*\equiv \cJ_{2,\xi}^*+\lim_{\epsilon\ra 0}\cJ_{2,\epsilon}^*=
\beta^{\frac{1}{2}} \bigg[\frac{1}{\xi\beta}-\frac{2}{\sqrt{\xi \beta}}
+\pi  \, y \ln\left(4 \pi e^{-\gamma}  \beta \xi\right)
+\frac{\pi}{3}\bigg],\qquad y\ra 0, \, \xi\ra 0
\end{equation}
This result holds if  $y$ and $\xi$ respect additional conditions,
converted into constraints on the mass scales in Appendix D. 
Eqs.(\ref{cj2star}), (\ref{j2starxi}) give $\cJ_2$ used in the text
eq.(\ref{dg2}). For comparison see eqs.(A-12) to (A-19) in
ref.\cite{Ghilencea:2002ff}.

%%%%%%%%%%%%%%%%%%%%%%%%%%%%%%%%%%%%%%%%%%%%%%%%%%%%%%%%
%% J3
%%%%%%%%%%%%%%%%%%%%%%%%%%%%%%%%%%%%%%%%%%%%%%%%%%%%%%%%

\newpage
\setcounter{equation}{0}
\def\theequation{C-\arabic{equation}} 
\subsection*{C. Kaluza-Klein Integrals: $\cJ_3$}
\noindent
$\bullet$ We  introduce $\cJ_3^*$  ($y>0$)
\begin{equation}\label{cj3star}
\cJ_3^*(\beta, y)=\sum_{m_1}'\sum_{\tilde m_2}'
\frac{1}{\sqrt\beta}\int_{\xi}^{\infty}\frac{dt}{t^{3/2}}
e^{-\frac{\pi \tilde m_2^2}{\beta \, t}-\pi \beta \, t (U_2^2 m_1^2+y)}\,
e^{-2\pi i \tilde m_2 U_1 m_1}
\end{equation}
and evaluate it in the limit $\xi\ra 0$, $y \ra 0$. \\
\noindent
For $\beta=1/(T_2 U_2)$ and $y=\chi/\beta$
%% $y=Y (T_2 U_2)$
one recovers   the integral of $\cJ_3$  of eq.(\ref{dg3}) in the text.\\

\noindent
 $\cJ_3^*$  is well defined and finite in  the UV limit, $t\ra 0$
due to the  non-vanishing exponentially suppressed term
$exp(-\tilde m_2^2/t)$.  Thus no UV divergence (in $\xi$) 
can appear and one can then  set $\xi=0$. 
To evaluate  the error introduced while 
doing so, see eq.(\ref{delt3}).  Note that 
$\cJ_3^*$ does not have  divergences in $y\!\ra \! 0$ (we assume
$U_2\not=0$), see eq.(14) of \cite{Ghilencea:2002ff} where 
$y=0$ case is evaluated; thus terms divergent in $y$ e.g. 
$\xi/y$ or $\xi\ln y$ are not present or ``missed''  when we  set $\xi=0$). 

 Using the integral representation 
of Bessel functions, eq.(\ref{bessel1})  we find 
\begin{eqnarray}\label{udependence}
\cJ_3^*&=&\sum_{m_1=1}^{\infty}\sum_{\tilde
m_2=1}^{\infty}\frac{2}{\tilde m_2} e^{-2\pi \tilde m_2 (y+m_1^2
U_2^2)^{1/2}}\, e^{2\pi i U_1 \tilde m_2 m_1}+c.c.\nonumber\\
&=&-2\ln \prod_{m_1\geq 1}\bigg\vert 1-e^{-2\pi (y+m_1^2
U_2^2)^{1/2}}\,  e^{2\pi i U_1 m_1}\bigg\vert^2
\end{eqnarray}
The product above may further be written as $\prod
(1+a_{m_1})$ and converges uniformly since the associated 
series $\sum \vert a_{m_1}\vert $ converges uniformly. One can therefore take 
the limit $y\!\ra\! 0$  on $\cJ_3^*$ to find
\begin{equation}\label{udpd}
\cJ_3^*=-2 \ln \prod_{m_1\geq 1}\bigg\vert 1-e^{2\pi i m_1 U}\bigg\vert^2
\end{equation}
Eq.(\ref{cj3star}) and (\ref{udpd}) give $\cJ_3$ in the text
eq.(\ref{dg3}) for $\chi \ll \beta$. \\

%%%%%%%%%%%%%%%%%%%%%%%%%%%%%%%%%%%%%%%%%%%%%%%%%%%%%%%%%%%%%%%%%%%%%%%%

%%  corrections

%%%%%%%%%%%%%%%%%%%%%%%%%%%%%%%%%%%%%%%%%%%%%%%%%%%%%%%%%%%%%%%%%%%%%%%%

% \newpage
 \vspace{1.5cm}
 \setcounter{equation}{0}
 \def\theequation{D-\arabic{equation}} 
 \subsection*{D. Vanishing errors in the regularised field theory
 result.}
%\label{apd}
 \noindent
 While computing $\cJ_1$ of eq.(\ref{dg1}), $\cJ_2$ of eq.(\ref{dg2}) 
and $\cJ_3$ of eq.(\ref{dg3}) we introduced additional errors for 
each of these. We ensure these errors  are negligible 
for ``small enough'' $\xi$ and $\chi\equiv \beta y$ and would like 
to convert  this condition into  bounds on the
 mass scales of the theory.\\

\noindent
$\bullet$ 
For $\cJ_1$ an error $\delta_1$ 
 arises in the step from  (\ref{sl1}) to (\ref{sl2});
for $\cJ_2$ an error $\delta_2$
arises in the step from (\ref{j2s2}) to (\ref{j2star});
for $\cJ_3$ an error $\delta_3$ arises from 
(\ref{cj3star}) (by setting $\xi=0$). 
Their  expressions are 
\begin{eqnarray}\label{delt1}
\delta_1& \equiv &\frac{1}{\sqrt\beta}\int_{0}^{\xi} \frac{dt}{t^{3/2}}
\sum'_{\tilde m_2} e^{-\pi \tilde m_2^2/(\beta t) -\pi \beta t y }, \qquad
\qquad \beta= \frac{1}{T_2 U_2}, \, \, y = \chi (T_2 U_2)
\\
\delta_2& \equiv &\frac{\sqrt{T_2 U_2}}{\sqrt \beta} \int_{1/\xi}^{\infty} dt \sum'_{\tilde
m_1} e^{-\pi t {\tilde m_1^2}/{\beta}-\pi  \beta y/t},\qquad\quad
 \beta=\frac{U_2}{T_2}, \, \,\,\,\,\,\,\,\, y=\chi \frac{T_2}{U_2}\label{delt2}
\\
\delta_3& \equiv & \frac{1}{\sqrt{\beta}}
\int_{0}^{\xi}\frac{dt}{t^{3/2}}\sum'_{m_1}\sum'_{\tilde m_2} e^{-\pi
 \tilde m_2^2/(\beta t)-\pi
\beta t (m_1^2 U_2^2+y)-2\pi i U_1 \tilde m_2  m_1},\,\,\,
 \beta=\frac{1}{T_2 U_2}, \, y= \chi (T_2 U_2)\label{delt3}
\end{eqnarray}
This gives, for any positive $y$
\begin{eqnarray}
\vert\delta_1\vert &\leq  & 
\int_{T_2 U_2/\xi}^{\infty} \frac{dt}{t^{1/2}}
\sum'_{\tilde m_2} e^{-\pi \tilde m_2^2 t}\\
\vert\delta_2\vert &\leq &  U_2 \int_{0}^{\xi U_2/T_2} \frac{dt}{t^2}
 \sum'_{\tilde
m_1} e^{-\pi {\tilde m_1^2}/t}\\
\vert\delta_3\vert & \leq & \sqrt{T_2 U_2}
\int_{0}^{\xi}\frac{dt}{t^{3/2}}\sum'_{m_1}\sum'_{\tilde m_2} e^{-\pi
 \tilde m_2^2 T_2 U_2/t-\pi
t m_1^2 U_2/T_2} 
\end{eqnarray}
A sufficient condition for the $\delta_i$ to vanish is 
(using eqs.(A-20) to (A-34) of 
Appendix A.1 of ref. \cite{Ghilencea:2002ff})
\begin{equation}
\frac{T_2 U_2}{\xi}\gg \textrm{max} \left\{U_2^2,
\frac{1}{U_2^2}\right\}
\end{equation}
With the notation $\Lambda^2 \equiv \mu^2/\xi $ this means
\begin{eqnarray}\label{rmreg1}
&&\Lambda \,R_2 \sin\theta > \Lambda \, R_1  \gg 1, \qquad\qquad
\,\,\, (U_2>1), \nonumber\\
&&\Lambda\, R_1 > \Lambda \, R_2 \sin \theta \gg (\Lambda
R_1)^{\frac{1}{2}},  \qquad (U_2<1),
\end{eqnarray}
used in the text, eq(\ref{cutoff}).
These also lead to a large compactification area
 $T_2^*\equiv \gL^2 R_1 R_2 \sin \gth \gg 1$. \\

%\vspace{0.2cm}
\noindent
$\bullet$ 
The limit of ``removing'' the infrared regulator $\chi\equiv
y\beta$,  provides further bounds on the mass scales of the theory:

\noindent
{\bf{(a)}.} While computing $\cJ_1$ in the text using $\cJ_1^*$ of 
 eq.(\ref{cj1star}) in the step  from eq.(\ref{j1star}) to (\ref{finalj1star})
a (vanishing) correction (in $\xi$ and $\chi\equiv \beta y$) is discarded
when the regulators are removed.  This  is converted into the following bounds
on the mass scale $Q^2\equiv \pi e^\gamma \chi \mu^2$
 associated with the infrared  regulator   $\chi$
\begin{equation}\label{rmreg2}
 Q^2\ll e^\gamma \Lambda^2,\qquad \quad  Q^2\ll 
\frac{e^\gamma}{4}\frac{1}{(R_2\sin\theta)^2}\\
\end{equation}
{\bf (b).} In computing $\cJ_2$ in the text eq.(\ref{dg2}) 
using $\cJ_2^*$ of
 eq.(\ref{cj2star})  in the steps
 from eq.(\ref{cj2starxi0}) to eq.(\ref{cj2starxi}) and from 
eq.(\ref{finalcj2starep}) to eq.(\ref{B16}) respectively,
we discarded additional corrections (in $\xi$ and $y\equiv\chi/\beta$)
while removing the regulators.
Ensuring that  these corrections are negligible provides 
the following (sufficient) conditions for the mass scales of 
the theory
\begin{eqnarray}\label{rmreg3}
&&Q^2\ll  e^\gamma\Lambda^2,\qquad\quad Q^2\ll e^\gamma \frac{1}{R_1 R_2
\sin\theta},\nonumber\\
&& Q^2\ll e^\gamma \mu^2,\qquad \quad Q^2\ll  \pi e^{\gamma}\frac{1}{R_1^2}
\end{eqnarray}
Eqs.(\ref{rmreg1}), (\ref{rmreg2}), (\ref{rmreg3}) combined together lead to
eq.(\ref{cutoff}) up to (regularisation dependent) 
coefficients of order unity, not shown there explicitly.

\end{document}